\documentclass[%
 reprint,
superscriptaddress,
 showkeys,
 nofootinbib,
 nobibnotes,
 amsmath,amssymb,
 aps,
 pra,
 floatfix,
]{revtex4-1}

\usepackage{graphicx}
\usepackage{grffile}

\usepackage{dcolumn}
\usepackage{bm}
\usepackage{hyperref}
\usepackage{subfigure}
\usepackage{float}

\usepackage[english]{babel}
\usepackage[utf8]{inputenc}
\usepackage[T1]{fontenc}
\usepackage{xspace}

\usepackage{xcolor}

\usepackage{ulem}

\newcounter{maintextfigures}

\begin{document}


\title{Scattering from controlled defects in woodpile photonic crystals}

\author{Stefan \surname{Aeby}}
\affiliation{%
Department of Physics, University of Fribourg, 1700 Fribourg, Switzerland 
}%
\author{Geoffroy J. \surname{Aubry}}%
\email{geoffroy.aubry@unifr.ch}
\affiliation{%
Department of Physics, University of Fribourg, 1700 Fribourg, Switzerland 
}%
\author{Nicolas \surname{Muller}}
\affiliation{%
Department of Physics, University of Fribourg, 1700 Fribourg, Switzerland 
}%
\affiliation{%
HES-SO,  Fribourg, Switzerland 
}%
\author{Frank \surname{Scheffold}}%
\email{frank.scheffold@unifr.ch}
\affiliation{%
Department of Physics, University of Fribourg, 1700 Fribourg, Switzerland 
}%

\date{\today}

\begin{abstract}
Photonic crystals with a sufficiently high refractive index contrast display partial or full band gaps. However, imperfections in the metamaterial cause light scattering and extinction of the interfering propagating waves. Positive as well as negative defect volumes may contribute to this kind of optical perturbation. In this study, we fabricate and characterize three-dimensional woodpile photonic crystals, with a pseudo-bandgap for near-infrared optical wavelengths. By direct laser writing, we intentionally introduce random defects in the periodic structure. We show that we can model random defect scattering by considering the difference between the disordered and the regular structure. Our findings pave the way towards better control and understanding of the role of defects in photonic materials that will be crucial for their usability in potential applications. 
\end{abstract}

\maketitle

\normalem

\section*{Introduction}

Dielectric materials with a periodic variation of the refractive index display photonic stop bands for an optical wave propagating in specific directions~\cite{Joannopoulos2008}.
The Bragg length $L_\mathrm{B}$ is a measure for the number of crystal layers penetrated by the incident beam and it is directly related to the scattering strength of the crystal layers. The attenuation of the wave  in the direction of a stopband scales roughly exponentially as the thickness $L$ of the crystal is increased beyond  $L_\mathrm{B}$~\cite{Galisteo-Lopez2003,Marichy2016}.
 For a perfect crystal, and $L \to \infty$, destructive interference always leads to vanishing transmission in the stop band's direction irrespective of the refractive index contrast~\cite{Spry1986}. Bandgaps in one and two-dimensional photonic crystals are now widely employed in applications such as supercontinuum fiber lasers~\cite{Russell2003} and for data processing using optical modules based on 2D silicon photonic crystal technology~\cite{Hochberg2010}.
 \newline  In a three-dimensional (3D) full photonic bandgap (PGB) material, for specific wavelengths, the propagation of light is inhibited in all directions. However, only above a certain threshold refractive index contrast full photonic bandgaps exist~\cite{Joannopoulos2008,Ho1994}.  3D photonic crystals (PCs) are metamaterials fabricated from mesoscopic building blocks. In contrast to the case of atomic or molecular crystals, these building blocks are not identical. Therefore, all photonic crystal materials possess some intrinsic degree of disorder due to surface roughness, size dispersion, stress-induced deformations, in addition to defects, stacking faults, crystal grain boundaries. The influence of disorder is often so strong that the interaction of the propagating wave with the periodic Bragg planes competes with scattering on similar length scales. In early work on opal photonic crystals the manifestation of intrinsic defects has been discussed and a plethora of studies found that it is nearly impossible to fabricate perfect crystals made by self-assembly of colloids \cite{Vlasov2000}. As a consequence, artificial opals of polystyrene spheres or air holes in TiO$_2$ (titanium dioxide) display diffuse, multiple-scattering in tandem with Bragg diffraction~\cite{Koenderink2003,Huang2001}.
Similar observations have been made for quasi-crystals~\cite{Ledermann2006}.
Improved self-assembly protocols and lithography have led to higher quality photonic crystals, but despite the progress made, imperfections still play a significant role~\cite{Lopez2003,Soukoulis2011}.
While disorder in photonic crystals is often considered a nuisance, it also highlights the rich and fascinating interplay between defect states, wave tunneling and percolation, random diffuse scattering, and directed Bragg scattering of light ~\cite{Skipetrov2004,Florescu2010,Froufe-Perez2017,Fernandes2013,Pratesi2013,Aubry2020}. Moreover, the interaction between the band structures and defect scattering might facilitate the observation of other critical coherent transport phenomena such as Anderson localization of light. In a seminal paper published in 1987, Sajeev John suggested the presence of localized defect states close to the band edge of a photonic crystal, due to multiple scattering and a reduced density of states \cite{John1987}. Finally, defect states can be introduced in a photonic crystal deliberately to implement a particular function, such as for optical sensing applications, lasing, or optical circuitry ~\cite{Joannopoulos2008,Ishizaki2013}. Controlling and understanding the role of defects and disorder in photonic crystals is thus of paramount importance.

\section*{Results}
\begin{figure}
    \includegraphics[width=\columnwidth]{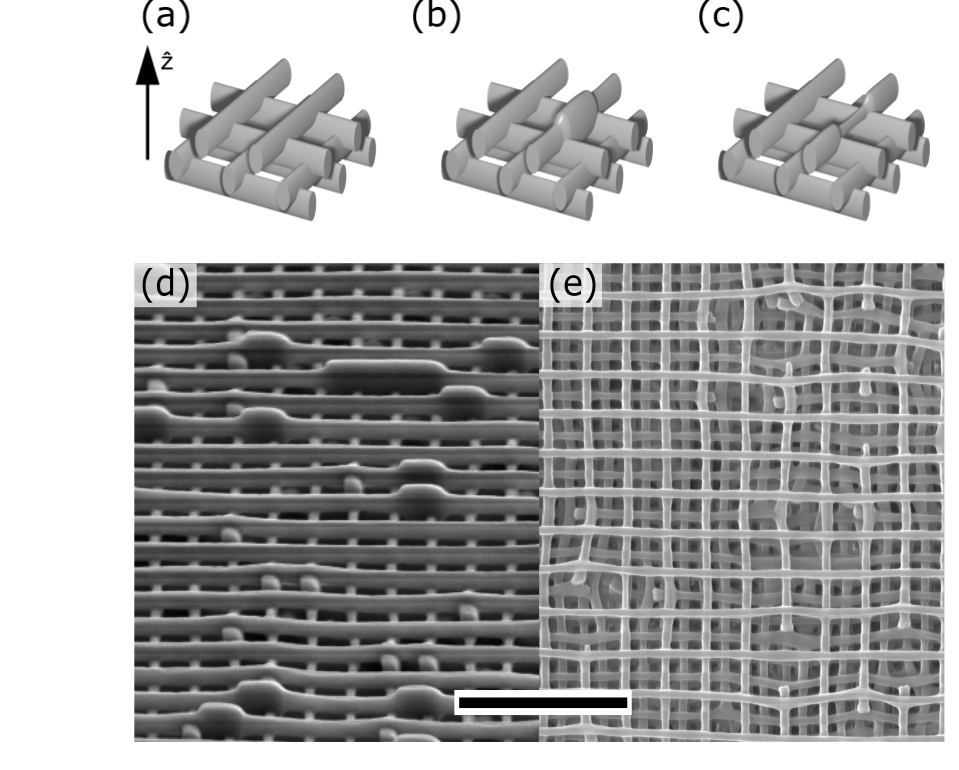}
    \caption{Computer generated renderings of (a) woodpile structure, (b) a woodpile structure with positive defects, (c) a woodpile structure with negative defects.
    Scanning electron micrograph of polymer woodpile structures fabricated by DLW, with (d) positive (rod cross section area +280\%, image taken at an angle of 52$^\circ$), and (e) negative defects (-100\%, top view).
    Scale bar is 5\ $\mu$m.
    }
    \label{fig:woodpile}
\end{figure}

\begin{figure}
    \includegraphics[width=\columnwidth]{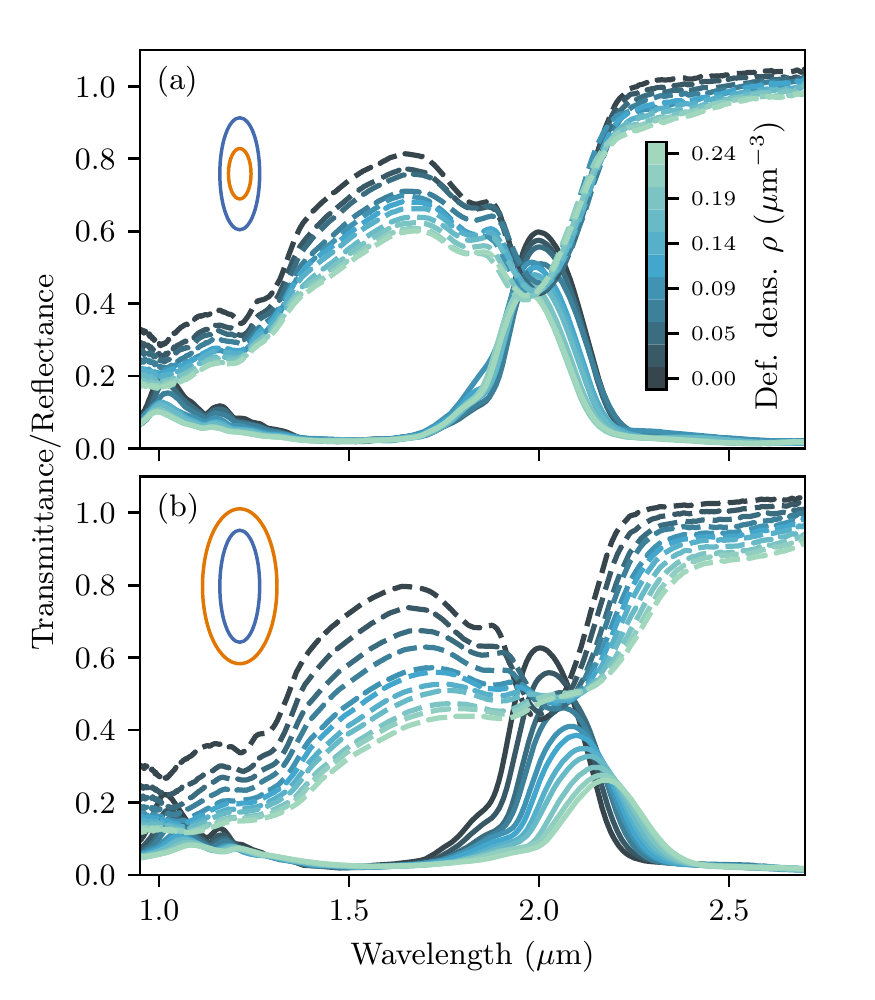}
    \caption{Experimental reflectance (solid lines) and transmittance (dashed lines) of woodpile structures with different defect number densities. The density ranges from 0 to 0.24 $ \mathrm{\mu m^{-3}} $.
    (a) spectra for negative defects ($\kappa=-74\%$ cross section surface), (b) for positive ($\kappa=+158\%$).
    The blue ellipses in inset show the standard dimensions of the WP-rods, while the orange ellipses show the dimensions the negative (a) and positive (b) defects.}
    \label{fig:reflTrans}
\end{figure}

\begin{figure}
    \includegraphics[width=\columnwidth]{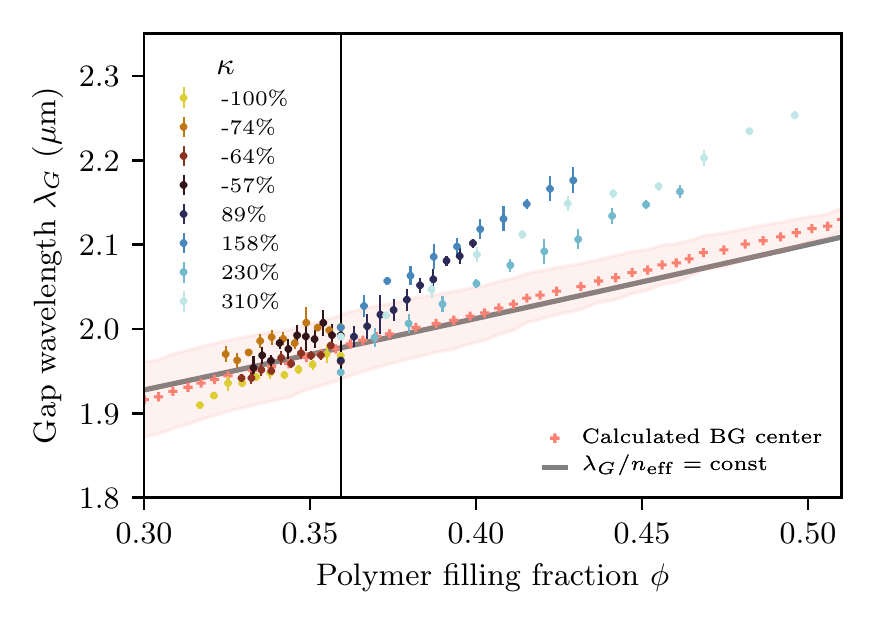}
    \caption{Center position (circles) of the reflection peak marking the position of the photonic pseudo-gap in the presence of defects.
    Each color corresponds to a different defect size.
    The values of $\kappa$ denote the volume of the defect in units of the unperturbed rod-segment. $-100$\% indicates a missing link, $+310$\% means the interstitial space between parallel rods is entirely filled, marking the two extreme cases.
    The error bars correspond to the standard deviation of the different realizations for equal fabrication parameters.
    The vertical line indicates the filling fraction of the  WP crystal.
    In red: pseudo-gap span (area) and center (crosses) calculated with MPB for the WP crystal with elliptical rods as a function of the rods filling fraction keeping the rods aspect ratio constant.
    The grey line shows the expected scaling of the gap wavelength assuming a linear dependency with the Maxwell-Garnet effective refractive index $n_\text{eff}$.
    }
    \label{fig:fillfrac}
\end{figure}

\begin{figure}
    \includegraphics[width=\columnwidth]{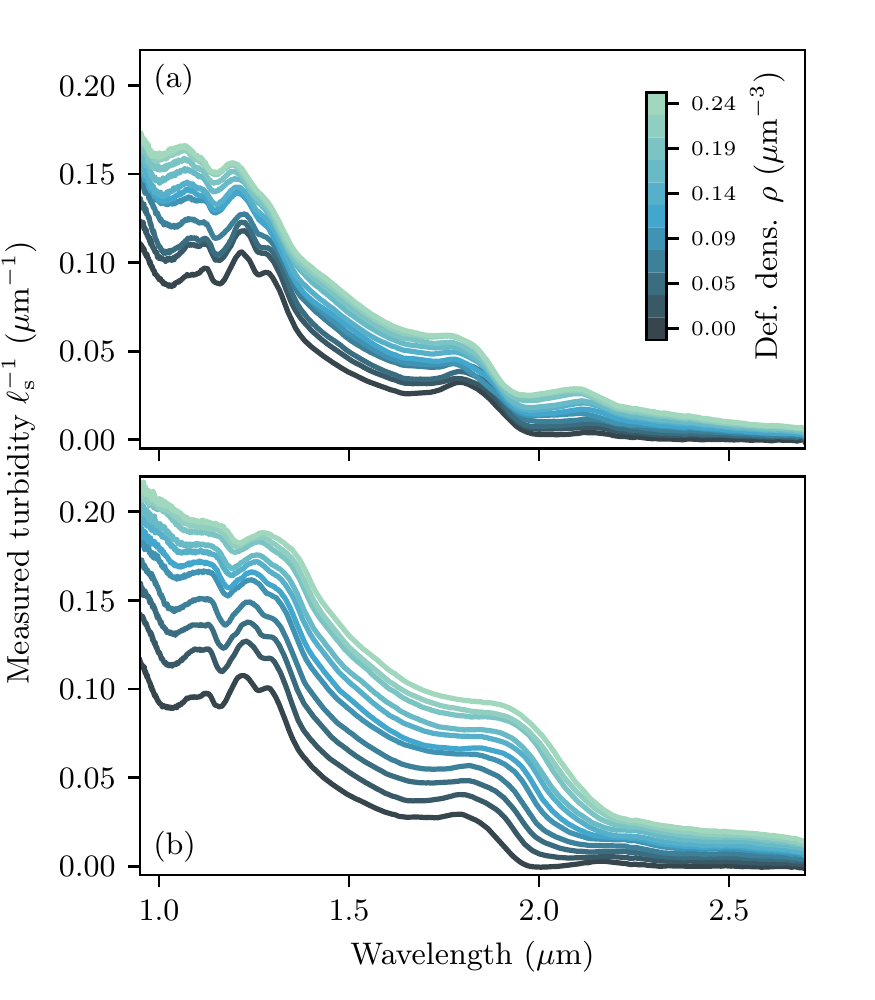}
    \caption{Turbidity (inverse of the scattering mean free path) of woodpile structures with (a) negative and (b) positive defects.
    The different color of the curves encode the defect number density $\rho$ in $\mathrm{\mu m^{-3}} $ in the sample. The samples are the same as in Fig.~\ref{fig:reflTrans}.
    }
    \label{fig:ls}
\end{figure}

Here, we report on a study about intrinsic and intentionally added defects in photonic crystals (PCs)~\cite{Yablonovitch1987}. The tight control over the position and size of the defects we have, sets our study apart from earlier work on intrinsic defects ~\cite{Koenderink2005} or intentionally added defects in opals~\cite{Garcia2011}. 
By direct laser writing (DLW) in a polymer resist~\cite{Sun1999,Deubel2004} (Photonic Professional, Nanoscribe, Germany), we fabricate high-quality woodpile (WP) photonic crystals.
For optimal results, we use the IP-Dip photoresist (Nanoscribe, Germany), refractive index $n_\mathrm{IP-Dip}=1.53$~\cite{Dottermusch2019}.
Our PCs display a pseudo-gap in the near-infrared range as shown in Fig.~\ref{fig:woodpile}(a)~\cite{Ledermann2006}. We intentionally add defects to our crystals to probe the effect of disorder on the photonic properties of the PCs. We introduce two types of defects: positive and negative defects, see Fig~\ref{fig:woodpile}, and we place these defects uniformly over the sample volume with a variable defect number density $\rho$. 
\newline The basic WP-structures are composed of horizontal arrays of parallel rods, where $d$ denotes the in-plane distance between two rods. Alternating layers are rotated by 90$^\circ$ and shifted by $d/2$ every 2 layers.
Hence, the structure repeats itself every 4 layers in the stacking direction given be $d_{z}$.
We use a ratio $d_{z}/d=\sqrt{2}$ which results in a face-centered-cubic-lattice (FCC, see Supplementary Information Fig.~\ref{fig:woodpileFCC}).
The rods of adjacent layers slightly overlap, as indicated in Figure \ref{fig:woodpile} (a-c), and we define a \emph{rod-segment} as the distance between two rod intersections. Note that one rod-segment corresponds to the primitive unit cell lattice constant, see Fig.~\ref{fig:woodpileFCC}.
By locally enlarging (reducing) the thickness of a rod-segment, Fig.~\ref{fig:woodpile}(b,c), we create positive (negative) defects.
We denote the percent increase of the rod cross sectional area with $\kappa$.
We control the thickness of the rods by setting the power of the DLW-laser below or above the default power used to fabricate the woodpile crystal. For the crystal structures, the in-plane-distance between rods, which is equal to the rod-segment length, is set to $d=1.2\ \mu$m.
All the samples have a total size of $70 \times 70 \times 8.5\ \mu \mathrm{m}^3$ which corresponds to five lattice constant of the conventional cubic unit cell of the FCC lattice (20 rods layers) in $\hat{\mathbf{z}}$- direction, i.e., perpendicular to rods long axis, Fig.~\ref{fig:woodpile}~(a). 
\newline Due to the asymmetric shape of the DLW focal volume, set by the point-spread function of the DLW-microscope objective, the cross-section of the rods is elliptical with an aspect ratio of nearly three. For each laser power used, we measure the rod cross sectional area with a scanning electron microscope (SEM);
see Materials and Methods.
For the crystal, we find for the long axis $a=0.42\ \mu$m, and the short axis $b=0.15\ \mu$m.
\newline Figure~\ref{fig:reflTrans} shows the reflectance and transmittance spectra recorded for a series of samples.
The measurements were performed by Fourier Transform Infrared Spectroscopy (FTIR, Bruker Vertex 70, and Hyperion Spectrometer).
As we use a Cassegrain mirror-objective in the FTIR-microscope, the light is transmitted and detected along a hollow cone with an acceptance angle between $\theta_{\mathrm{min}}$=15$^\circ$ and $\theta_{\mathrm{max}}$=30$^\circ$ with $\theta_{\mathrm{mean}}$=22.5$^\circ$ compared to the $\Gamma-\mathrm{X}_z$-direction of the crystal (see Supplementary Information Fig.~\ref{fig:fccBrilloin} for the FCC-Brillouin zone scheme).
We measure the spectra averaged over an area of about 40x40\ $\mu \mathrm{m}^2$ covering wavelengths between 0.9 and $2.6\ \mu$m while for wavelengths larger than $2.6\ \mu$m, the polymer absorbs light (for details see 
Materials and Methods%
).
We use a silver-coated mirror as a reference for the reflectance spectra.
To calibrate the transmittance, we measure the spectrum of the bare substrate next to the samples.
For each set of fabrication parameters, we produce several samples, and each curve in Fig.~\ref{fig:reflTrans} represents an average over three to five different sample realizations.
The darkest lines shows the spectra of the crystalline samples, i.e., without any added defects. The transmittance (reflectance) displays a profound dip (peak) indicating the presence of a pseudo-gap in the chosen incident direction. We find that by increasing the number density $\rho$ of  defects, the peak of the reflectance and the dip of the transmittance become less marked meaning that the quality of the bandgap is affected by the presence of defects. Interestingly, we also observe that the peaks are shifted to higher (lower) wavelengths when increasing the density of positive (negative) defects.
To study this effect quantitatively, we plot in Fig.~\ref{fig:fillfrac} the position of the reflectance maxima for different defect sizes and defect densities against the polymer filling fraction $\phi$ of each sample (see 
Materials and Methods
for the method used to estimate the filling fraction).
Interestingly, we find that all the data collapse on a master curve. The pseudo-gap center wavelength increases roughly linearly with $\phi$, independent of whether the defect volume or number density is varied to achieve a certain $\phi$-value. Our results are consistent with effective medium theories, such as the Maxwell-Garnett mixing formula~\cite{Garnett1904}, that predict that the effective refractive index $n_\mathrm{eff}$ roughly scales linearly with $\phi$ in the studied range, see the yellow line in Fig.~\ref{fig:fillfrac}. This shift of $\lambda_G$ can be explained as follows. The reflectance maximum is due to Bragg back-scattering at a scattering angle $\Theta \simeq 180^\circ$ where the momentum transfer $\vec{q}=2\vec{k}\sin[\Theta/2]\equiv \vec{G}$ matches a reciprocal lattice vector $\vec{G}$ with $\left |k \right |=2\pi n_\text{eff}/\lambda_G$. Thus, we expect that the maximas' $\lambda_G$ scale with the effective refractive index of the medium $n_\text{eff}$ such that $\lambda_G/ n_\text{eff}$ remains constant. We, therefore, can explain the linear increase of the bandgap-center wavelength with a change in the average polymer filling fraction, and we find this effect to be surprisingly robust against defect scattering.
\newline We corroborate this finding's accuracy by band structure calculations with MPB (MIT Photonic Bands)~\cite{mpb} for WP photonic crystal with different filling fractions. For all the filling fractions, the calculations, shown as red crosses in Fig.~\ref{fig:fillfrac}, were performed for elliptical rods with a constant aspect ratio of $a/b=0.42/0.15=2.8$. To accurately compare the simulations with the measurements using a Cassegrain-objective, we proceed as follows. We calculate the intersections of the widths of the gaps for all the directions, making an angle $\theta_{\mathrm{eff}}$ relative to the $\Gamma-\mathrm{X}_z$ direction where $\theta_{\mathrm{eff}}$ is defined by the incident direction and the refraction at the interface,
for details see 
Materials and Methods.
With increasing filling fraction $\phi$, the MPB-value of $\lambda_G$ increases and its evolution follows the Maxwell-Garnett effective refractive index, as shown in Fig.~\ref{fig:fillfrac}
\newline  Next, we study the scattering by the intentionally induced defects. We use a modified Beer-Lambert's law to extract the scattering mean free path $\ell_\mathrm{s}$ from the simultaneous reflection $R$ and transmission $T$ measurements of the PC,
\begin{align}
    T(L) + R(L) = \alpha\exp\left(-L/\ell_\mathrm{s}\right),
    \label{eq:ls}
\end{align}
where $L$ is the thickness of the sample.
This relation is valid before the onset of diffraction and for negligible absorption~\cite{Garcia2009}.
Both conditions are met in our case over the range of wavelength studied; see also Supplementary Information Fig.~\ref{fig:polymerAbsorption}.
The additional factor $\alpha$ in Eq.~(\ref{eq:ls}) takes account of systematic errors in the calibration procedure.
The bare substrate and the sample covered substrate deviate in two ways. First, the sample acts as an antireflection coating since the effective index $n_\text{eff}\simeq 1.3$ lies in between glass and air, which leads to an increased optical transmission $T$. Second, the sample thickness, $L\simeq8.5\ \mu$m, is only a small multiple of the wavelength, which results in Fabry-P\'erot interferences between the paths reflected by the sample top and the WP/substrate interface.  As a consequence, we observe slow oscillations in $T(\lambda)$ which complicates proper calibration. To achieve a model independent calibration, for each set of samples, we take averages of $T+R$ between $\lambda=2.45\ \mu$m, and $2.55\ \mu$m, where we expect $T+R\simeq \alpha$  for the crystal structure. We find typical values of $\alpha$ between 0.948 and 0.964. From the data shown in Fig.~\ref{fig:reflTrans}, and using Eq.~(\ref{eq:ls}), we extract $\ell_\mathrm{s}$ as a function of $\lambda$ as shown in Fig.~\ref{fig:ls}. 

\section*{Discussion}

\begin{figure}
    \centering
    \includegraphics[width=\columnwidth]{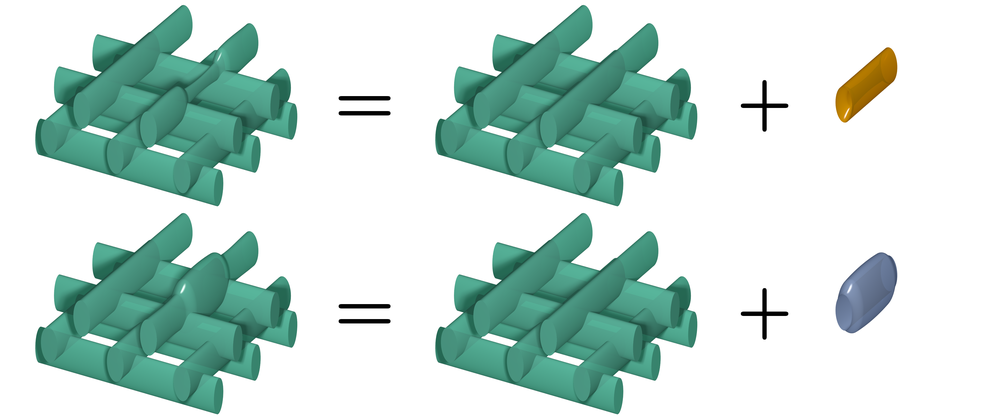}\\
    \includegraphics[width=0.7\columnwidth]{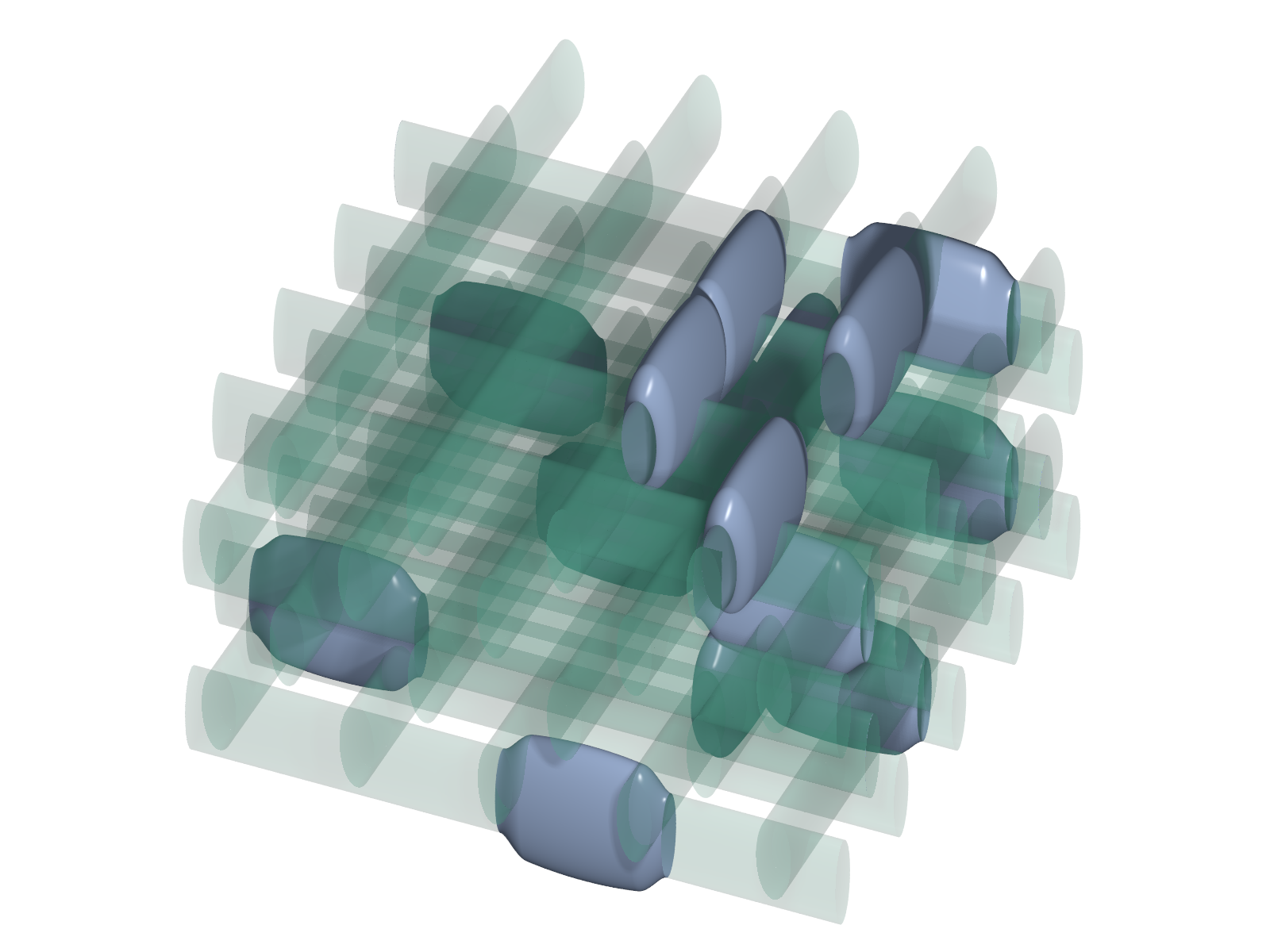}
    \caption{The measured turbidity of the samples is the sum of two contributions, the intrinsic turbidity of the woodpile crystals (due to DLW imperfections) on one hand, and the induced turbidity due to the artificially added defects on the other hand.}
    \label{fig:defectsRGD}
\end{figure}

\begin{figure}
    \includegraphics[width=1\columnwidth]{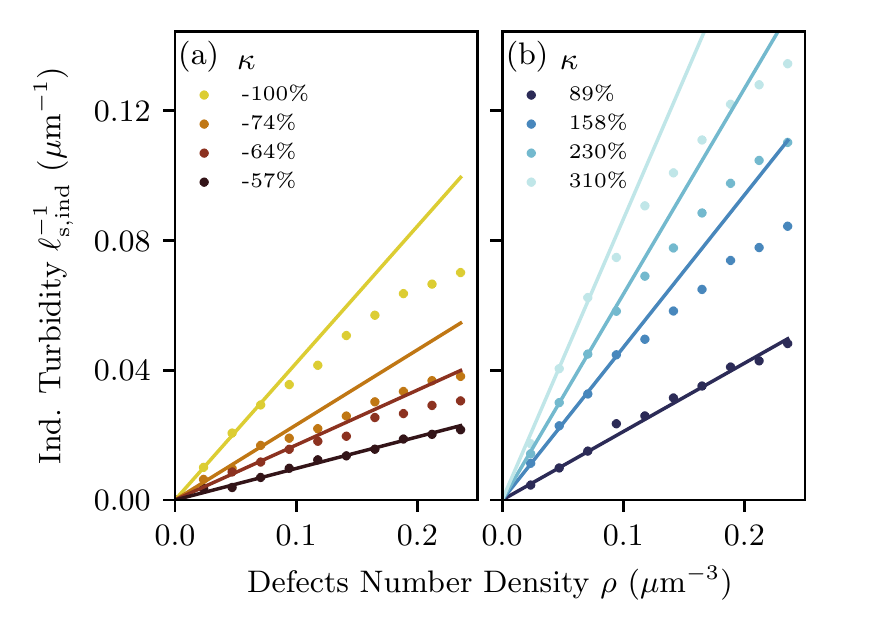}
    \caption{Induced turbidity for different defect volumes measured at $\lambda_0=1.57\ \mu$m for negative (a) and positive defects (b) as a function if the defects number density.
    Each color corresponds to a different relative change of the rod cross section for a single defect with respect to the default rod. Lines are linear fits done on the first points.}
    \label{fig:lsSingleWavelength}
\end{figure}

We first note that the turbidity of the WP crystals, given by the reciprocal of the scattering mean free path, $1/\ell_\mathrm{s}$, is finite, and it rises in the low-wavelength regime. The residual scattering from WP crystals, observed previously in~\cite{Deubel2004}, is caused by intrinsic roughness, small displacements, undulations, and deformations in the crystal. The turbidity of a crystal in the presence of artificially added defects is therefore determined by defect scattering and by intrinsic scattering.
In Fig.~\ref{fig:defectsRGD}, we illustrate how we can understand the scattering from our materials as a sum of scattering from the WPs and the intentionally added defects. 
If we assume that these two contributions contribute independently, we can express the measured turbidity $1/\ell_\mathrm{s}$ as follows~\cite{Pine1990}
\begin{align}
    \frac{1}{\ell_\mathrm{s}} = \frac{1}{\ell_\mathrm{s,0}} + \frac{1}{\ell_\mathrm{s,ind}},\label{uncorls}
\end{align}
where $1/\ell_\mathrm{s,0}$ corresponds to the intrinsic scattering and $1/\ell_\mathrm{s,ind}$ to the artificial defects scattering.
The same expression can be derived in the frame of the classical theory for the resistivity of ordinary metals~\cite{Ashcroft1976}. In the latter case the resistivity is controlled by the scattering mean free path of electrons;  $1/\ell_\mathrm{s,0}$ corresponds to the residual resistivity contribution, due to intrisic defects, and $1/\ell_\mathrm{s,ind}$ to the temperature dependent contribution $\propto T$, also known as the Wiedemann-Frantz law, due to electron-phonon scattering. 
\newline Both for negative (Fig.~\ref{fig:ls}(a)) and positive (Fig.~\ref{fig:ls}(b)) defects, we see that the induced turbidity increases with the defect number density $\rho$.  In Fig.~\ref{fig:lsSingleWavelength}, we plot the same data, $1/\ell_\mathrm{s,ind}$, at a selected wavelength ($\lambda_0 = 1.57\ \mu$m) as a function of the defect number density.
We plot the  data for negative and positive defects of different size in separate panels Fig.~\ref{fig:lsSingleWavelength} (a) and (b). The induced turbidity increases linearly with $\rho$ for sufficiently small defect number densities, supporting the validity of our assumptions in Eq.\eqref{uncorls} in this limit, $\rho\le0.1\ \mu$m$^{-3}$. The linear scaling of $1/\ell_\mathrm{s,ind}$ suggests that induced defects act as independent scatterers. Indeed, the turbidity of a medium consisting of identical uncorrelated scatterers is given by $1/\ell_\mathrm{s} = \rho \sigma$, where $\sigma$ denotes the total scattering cross-section of a scatterer. For higher concentrations, positional correlations and proximity effects lead to deviations from the linear dependence~\cite{Fraden1990,RezvaniNaraghi2015,Aubry2017}.
\begin{figure}
    \centering
    \includegraphics[width=\columnwidth]{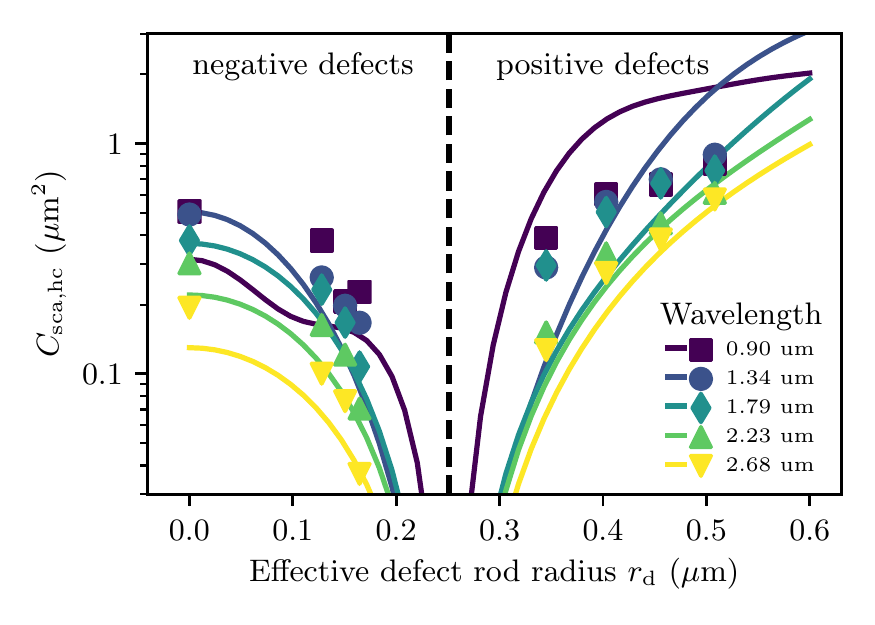}
    \caption{Total scattering cross-section of defect scatterers. Symbols show $C_\mathrm{sca,hc}$ extracted from the slopes of the curves $\left ( \rho \ell_\mathrm{s,ind}\right)^{-1}$ (solid lines shown in Fig.~\ref{fig:lsSingleWavelength}). The effective radius of the defect rod-segments is denoted by $r_\text{d}$ with $r_\text{d}=r=0.25 \mu$m for the standard rod size in the crystal (vertical dashed line).}
    \label{fig:Csca}
\end{figure}
Our results suggest that the initial slope of $1/\ell_\mathrm{s}(\rho)$ is a measure for the total scattering cross section of defects
\begin{align}
    \frac{1}{\ell_\mathrm{s,ind}} &= \rho C_\mathrm{sca,hc}.
\end{align}
In Fig.~\ref{fig:Csca}, we report the values for $C_\mathrm{sca,hc}$ extracted in the low-$\rho$ regime as a function of $r_\mathrm{d}=\sqrt{a_\mathrm{d}b_\mathrm{d}}$ where $a_\mathrm{d}$ and $b_\mathrm{d}$ are the long and short axis of the defect rod-segments. For all the samples studied, we report data for five different wavelengths. 
\newline Inspired by earlier work on disordered opals, Ref.~\cite{Koenderink2005}, we model the defect scattering in the frame of the 1\textsuperscript{st} Born, also known as Rayleigh-Gans-Debye (RGD) scattering, approximation. We hypothesize that we can treat the WP crystal as an effective, homogeneous background medium, and scattering arises from the local density differences, as shown in Figure~\ref{fig:defectsRGD}. To this end, we calculate the known RGD-expression for the total scattering cross-section of a hollow cylinder $C_\mathrm{sca,hc}$ with a length $d=1.2 \mu$m equal to a rod segment~\cite{Bohren1998}.

For simplicity, in our calculations, we consider scattering from hollow cylinders with $n_\mathrm{IP-Dip}$ in air ($n_\text{air}=1$), and limiting radii $r_\mathrm{d}=\sqrt{a_\mathrm{d}\cdot b_\mathrm{d}}$, $r=\sqrt{a\cdot b}=0.25\ \mu$m. For positive defects $r_\mathrm{d}>r$ while for negative defects $r_\mathrm{d}<r$. We explicitly take into account the optical geometry imposed by the Cassegrain objective. To this end, we calculate $C_\mathrm{sca,hc}$ for an angle between the incoming beam and the cylinder axis $\zeta=90^\circ-\left<\theta_\mathrm{eff}\right>=71.3^\circ$.
$\theta_\mathrm{eff}$ is the angle after refraction by the effective medium ($n_\mathrm{eff}$) when the incident angle is $22.5^\circ$ (Cassegrain objective).
We calculate $\left<\theta_\mathrm{eff}\right>$ by averaging $\theta_\mathrm{eff}$ over the different $n_\mathrm{eff}$ obtained by varying the filling fraction over the experimental range ($0.3 < \phi < 0.5$). 
The effective refractive index $n_\mathrm{eff}$
also enters via the effective wavenumber $k_\mathrm{eff} = 2\pi n_\mathrm{eff}/\lambda_0$.
\newline In contrast to earlier studies on disordered opals \cite{Koenderink2003,Koenderink2005}, our entire modeling predictions, shown as lines in Fig.~\ref{fig:Csca}, are fit-parameter free. Overall, we find an good agreement between the data and the model: the calculated scattering cross-sections follow the trend of the experimental data.
The agreement between theory and experiment is nearly quantitative for the higher wavelengths where the intrinsic scattering is negligible, and the transmission of the unperturbed crystal is high, signaling a trend toward an effective homogeneous medium. 
\newline In conclusion, in this study, we have quantified the effect of intrinsic and induced defects on photonic crystals' optical properties. We demonstrate that both the bandgap position and its quality are simultaneously affected by defect scattering.
The study presented in this work can provide essential guidelines how to quantify and model defect scattering. Our study also provides a modelling framework for diffuse scattering in PCs that lays the ground toward more complex disordered PCs based photonic materials. Such materials, based on higher refractive index building block, could become crucial to reach conditions for Anderson localization of light~\cite{Sperling2016}. 

\section*{Materials and methods}

\subsection*{Rod size measurement}
\label{sec:rodSize}
The lateral size of the rods, or the short axis of the elliptical cross section, can be assessed by taking a top-view electron micrograph. The measurement of the long axis of the elliptical rods inside the structure is more difficult. We tried to obtain a side-view of the rods using ion-beam milling but this posed problems due to the melting of the polymer when exposed to the ion beam.
To circumvent this problem, we wrote single rods with different laser powers between two large pillars, as shown Supplementary Information Fig.~\ref{fig:rodSize}.
This procedure allowed us to take SEM-images of the rods created by using different laser powers in DLW, from the top and at an inclined angle of 45$^\circ$.
We measure the short and long axis several times at different locations. The FWHM of the grey value profile of a line perpendicular to the rod is taken to determine the size.
We obtain a mean value by measuring the size on different positions on the same image and get a standard deviation of about 7\%, which we attribute to the uneven surface of the rods and the limited accuracy of the procedure.

\subsection*{Absorption spectrum of the IP-Dip polymer}

Supplementary Information Fig.~\ref{fig:polymerAbsorption} shows the absorption spectrum of a glass substrate coated with a $10\mu$m layer of developed IP-Dip photoresist. In the wavelength-band between $\lambda=1$ and $2.6\ \mu$m absorptive losses are small. Between $\lambda=2.45$ and $2.55\ \mu$m we observe a average drop of transmittance by 2.7\%. For the thinner ($L\simeq 8.5\mu$m) and air-filled WP-layers we estimate that absorptive losses attenuate the transmitted power by less than 1\%.

\subsection*{Assessment of the Polymer Filling Fraction }
\label{sec:fillingFraction}

The overlap between alternating layers of rods is taken into account in order to accurately estimate numerically the filling fraction of each sample.
To this end, we discretize the unit cell of the digital representation of the woodpile structure, and label every volume element (or voxel) belonging to at least one rod. 
We then calculate the ratio between the labeled and unlabeled voxels to get the filling fraction.
Using this method, we calculate $\phi_0$ for the unit cell of the perfect woodpile, but also $\phi_{0,\mathrm{d}}$ for the unit cell containing a single defect.
The parameters of the rods are taken from the rod size measurements.
For a sample having a defect density $\rho$, the filling fraction of the sample is then $\phi=(1-\rho V_0)\phi_0 + \rho V_0 \phi_{0,\mathrm{d}}$ where $V_0$ is the volume of the unit cell.

\subsection*{Band structure calculations}
\label{sec:mpb}

The woodpile is a FCC lattice, see Supplementary Information Fig.~\ref{fig:woodpileFCC}.
Its band structure is calculated in the first Brilloin zone, see Supplementary Information Fig.~\ref{fig:fccBrilloin}.
Usually, the (pseudo)gap is observed in the $\Gamma-\mathrm{X}_z$ direction (See Supplementary Information Fig.~\ref{fig:fccBrilloin}).
In our measurement, because of the Cassegrain objective, we observe transmission of light having an incident angle between 15$^\circ$ and 30$^\circ$.
We therefore have to compare our measurements with band structure calculation for the same angles.
To achieve this, we performed calculations of the band structure in the $\Gamma-\mathrm{P}_\theta(\varphi)$ directions as a function of the azimuthal angle $\varphi$ for a fixed polar angle of $\theta=\arcsin\left(\theta_\mathrm{mean}/n_\mathrm{eff}\right)$ with $\theta_\mathrm{mean}=(15^\circ+30^\circ)/2$ due to the Cassegrain objective and $n_\mathrm{eff}(\phi)$ calculated with the Maxwell-Garnett mixing formula~\cite{Garnett1904} (See Supplementary Information Fig.~\ref{fig:fccBrilloin} and its legend for the geometry).

Supplementary Information Figure~\ref{fig:bandStructure} shows the band structure for different $\Gamma-\mathrm{P}_\theta(\phi)$ directions.
$\theta$ is constant and $\phi$ is equally distributed between 0 and 90$^\circ$ (for symmetry reasons we don't need to calculate for all angles between 0 and 360$^\circ$ because the structure is invariant by a rotation of 90$^\circ$ around the $\hat{\mathbf{z}}$ axis).
The bandgap corresponds to the intersection of the bandgaps calculated for $\phi\in[0,90^\circ]$.

\subsection*{Rayleigh-Gans-Debye scattering of a hollow cylinder}
\label{sec:rgHollowCylinder}

The form factor of a finite cylinder of radius $r$ and length $L$ (see Supplementary Fig.~\ref{fig:cylinder} for the geometry), illuminated by a beam making an angle $\zeta$ with its axis oriented along the $\hat{\mathbf{z}}$ axis, as found in \citet{Bohren1998}, is
\begin{widetext}
\begin{align}
    f_\mathrm{c} &= \frac{1}{\pi r^2 L} \int_{-L/2}^{L/2}e^{-ikAz}dz\int_0^r\rho\, d\rho \int_0^{2\pi} e^{-ik\rho(B\cos \psi + C\sin \psi)} d\psi.
\end{align}
The integration is done in cylindrical coordinates $(\rho, \psi, z)$, which axis $\hat{\mathbf{z}}$ is oriented along the direction of the cylinder.
Note that in this calculation, to take advantage of the radial symmetry, the cylinders are oriented \emph{along} the $\hat{\mathbf{z}}$ axis, whereas in the rest of the paper they are \emph{perpendicular} to the $\hat{\mathbf{z}}$ axis.
$k=2\pi n_\mathrm{eff}/\lambda_0$ (where $n_\mathrm{eff}$ is the effective refractive index of the medium, and $\lambda_0$ the wavelength in vacuum of the incoming beam) is the wave number.
Finally,
\begin{align}
    A &= \cos \zeta + sin\theta \cos\varphi\\
    B &= \sin \theta \sin\varphi \\
    C &= \cos\theta -sin\zeta\\
    M &= \sqrt{B^2+C^2},
\end{align}
where $\theta \in \left[0,\pi\right]$ and $\varphi\in \left[0,2\pi\right]$ are the spherical coordinates angles for scattered wave.

To calculate the form factor of a hollow cylinder, one just has to perform the second integration between $r_1$ and $r_2$ (the inner and outer radii of the cylinder) instead of $0$ and $r$.
Performing all the integration, we obtain
\begin{align}
    f_\mathrm{hc}(\theta,\varphi;\zeta)=\frac{2}{L(r_2^2-r_1^2)}\frac{r_2J_1(kr_2M)-r_1J_1(kr_1M)}{kM}\frac{2\sin(kAL/2)}{kA}
\end{align}
$J_1$ is the Bessel function of the first kind of order 1.

The Rayleigh-Gans-Debye scattering cross section for a cylinder making an angle of $\zeta$ with respect to the incoming beam is then defined through the integral over all solid angles $\Omega$
\begin{align}
    C_\mathrm{sca,hc} (\zeta)&= \int_{4\pi}\frac{k^4}{4\pi^2}(m-1)^2 v^2 \left|f_\mathrm{hc}(\theta,\varphi; \zeta)\right|^2 \left[\cos^2\theta \cos^2\varphi + \sin^2\varphi\right] d\Omega.
    \label{eq:CscaHollowCylinder}
\end{align}
\end{widetext}
The integrations over $\theta$ and $\varphi$ are performed numerically.

In this paper, the incoming light makes an angle $\zeta=\frac{\pi}{2}-\theta_\mathrm{eff}$ with the axis of the cylinders, where $\theta_\mathrm{eff} = \arcsin{\left(\theta_\mathrm{mean}/n_\mathrm{eff}\right)}$.
$n_\mathrm{eff}$ should depend on the filling fraction, each curve and point plotted in the main text Fig.~\ref{fig:Csca} is deduced from measurements done varying the filling fraction.
We therefore kept $n_\mathrm{eff}$ constant and equal to the value of the Maxwell-Garnett refractive index of the perfect structure.

\begin{acknowledgments}
We thank Luis Froufe-P\'erez for fruitful discussions. We acknowledge financial support by the Swiss National Science Foundation under  grants No. 169074 and 188494. This work benefited from support from the Swiss National Science Foundation via the National Center of Competence in Research Bio-Inspired Materials. 
\end{acknowledgments}

\section*{Data and code availability}

All experimental and numerical data discussed in this manuscript, and the associated codes for analyzing or generating those data can be obtained upon reasonable request.

\section*{Competing interest}

The authors declare no competing interest.

\bibliography{sharedBiblio}

\clearpage

\setcounter{maintextfigures}{\value{figure}}
\renewcommand{\thefigure}{S\the\numexpr\value{figure}-\value{maintextfigures}\relax}
\setcounter{equation}{0}
\renewcommand{\theequation}{S\arabic{equation}}

\onecolumngrid
\section*{Supplementary Information}
\vspace{1cm}

This document contains all the figures referenced in the Materials and Methods section.

\begin{figure}[b]
    \centering
    \includegraphics{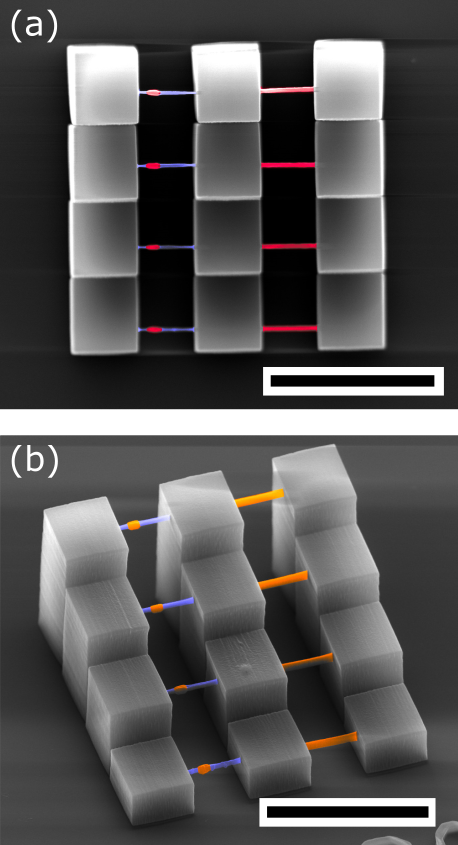}
   
    \caption{Structure printed by DLW to determine the dimensions of a single rod. (a) Top view; (b) same sample viewed from different angle.
    We placed the defects at different heights in order to allow a size measurement from a side view. This inforamtion is then used to calculate the long and short axis of the rod cross section as discussed in the text.
    We have color-code the image as follows: (blue) the laser power is set to the usual writing power of the woodpile crystal structure ($I_0$). On the left hand side of the structure, a single positive defect per line is shown in red ($2\, I_0$) (a) and orange ($1.44\, I_0$) (b). On the right hand side, the entire rod is written with the increased laser power.
    Scale bars are 20$\mu$m.}
    \label{fig:rodSize}
\end{figure}

\begin{figure}
    \centering
    \includegraphics{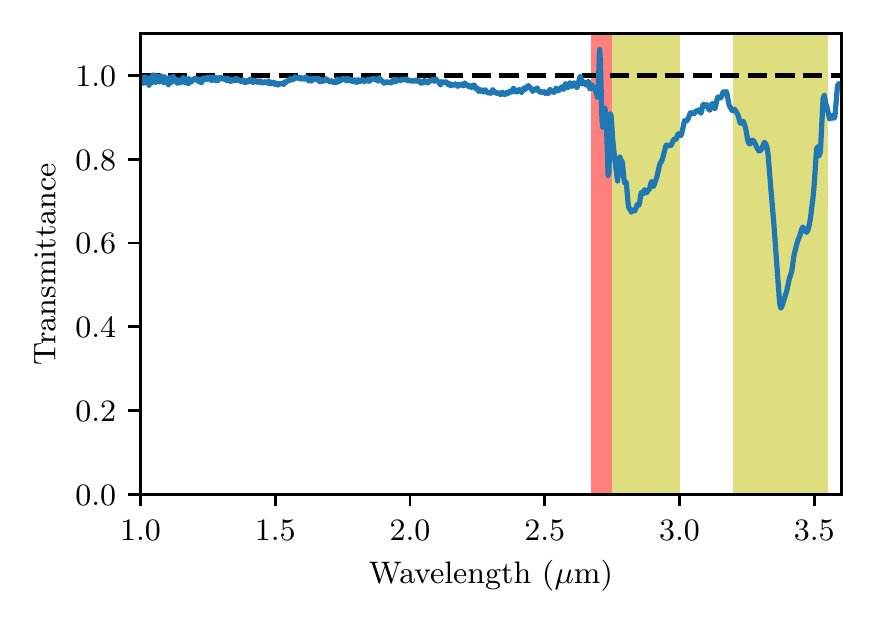}
    \caption{Absorption spectrum of a thin homogeneous film of made of the developed photoresin IP-Dip used in the study. The thickness of the layer is $10\ \mu$m.
    The red shaded area indicates the onset of absorption of the water vapour in air. The yellow shaded areas highlight the absorption bands of the polymer.
       }
    \label{fig:polymerAbsorption}
\end{figure}

\begin{figure}
    \centering
    \includegraphics[width=0.5\textwidth]{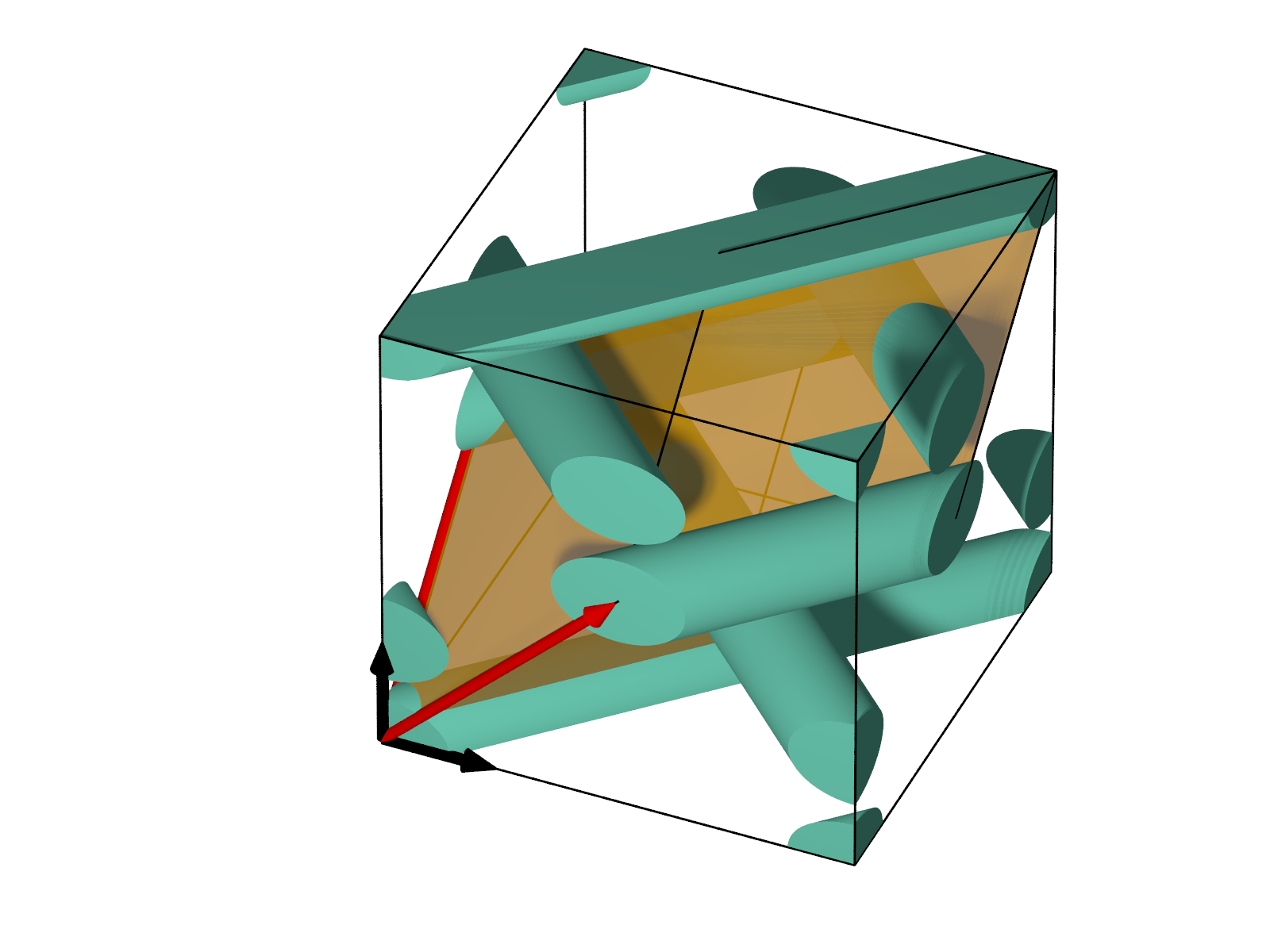}
    \caption{Conventional FCC unit cell of the woodpile structure.
    The black arrows correspond to the cartesians unit vectors, the red arrows to the FCC primitive vectors.
    The volume marked in dark yellow corresponds shows the primitive unit cell.}
    \label{fig:woodpileFCC}
\end{figure}

\begin{figure}
    \centering
    \includegraphics{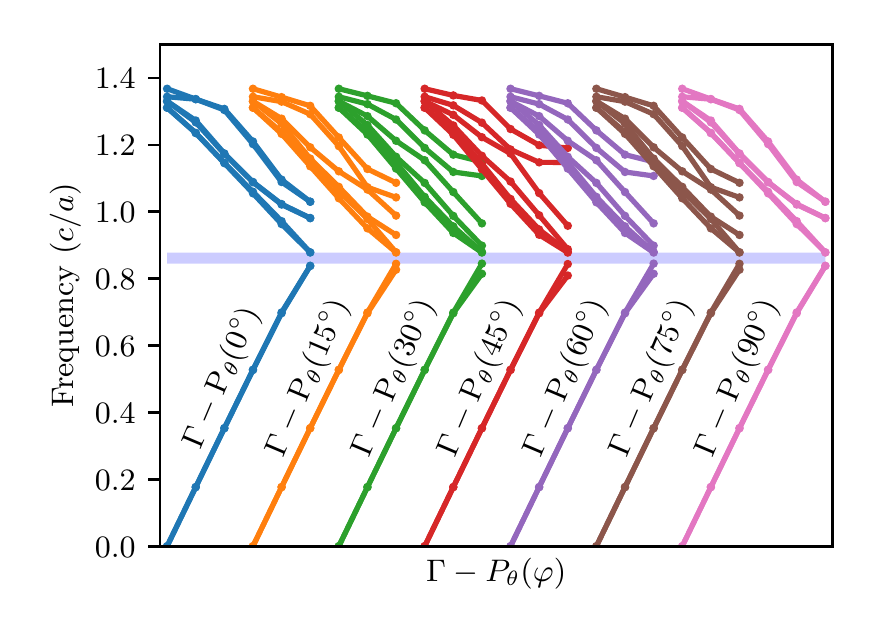}
    \caption{Band structure calculation for the directions making a polar angle $\theta=19.1^\circ$ with the $\hat{\mathbf{z}}$ direction calculated for the perfect structure where the cylinder cross-section have an aspect ratio $a/b$.
    The blue area is the bandgap.}
    \label{fig:bandStructure}
\end{figure}

\begin{figure*}
    \centering
    \includegraphics[width=\textwidth]{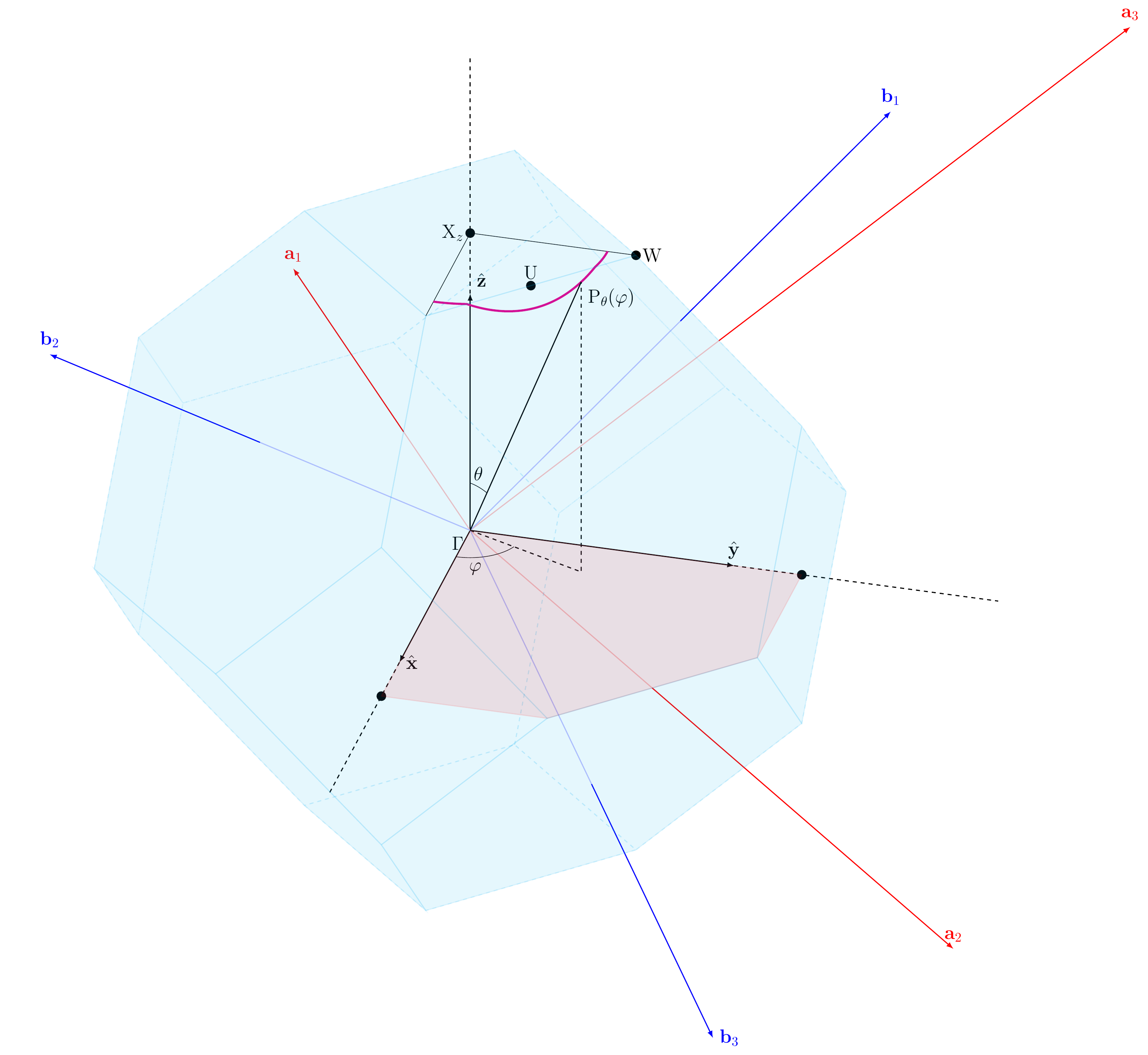}
    \caption{The blue volume corresponds to the first Brilloin zone of the FCC lattice whose primitive vectors are drawn in red.
    The blue vectors correspond to the reciprocal space primitive vectors.
    $\Gamma$, X$_z$, W, U are high symmetry points on the Brilloin zone surface.
    P$_\theta(\varphi)$ is the point on the first Broilloin zone whose coordinates are $(\theta,\varphi)$ in spherical coordinates.}
    \label{fig:fccBrilloin}
\end{figure*}

\begin{figure*}
    \centering
    \includegraphics{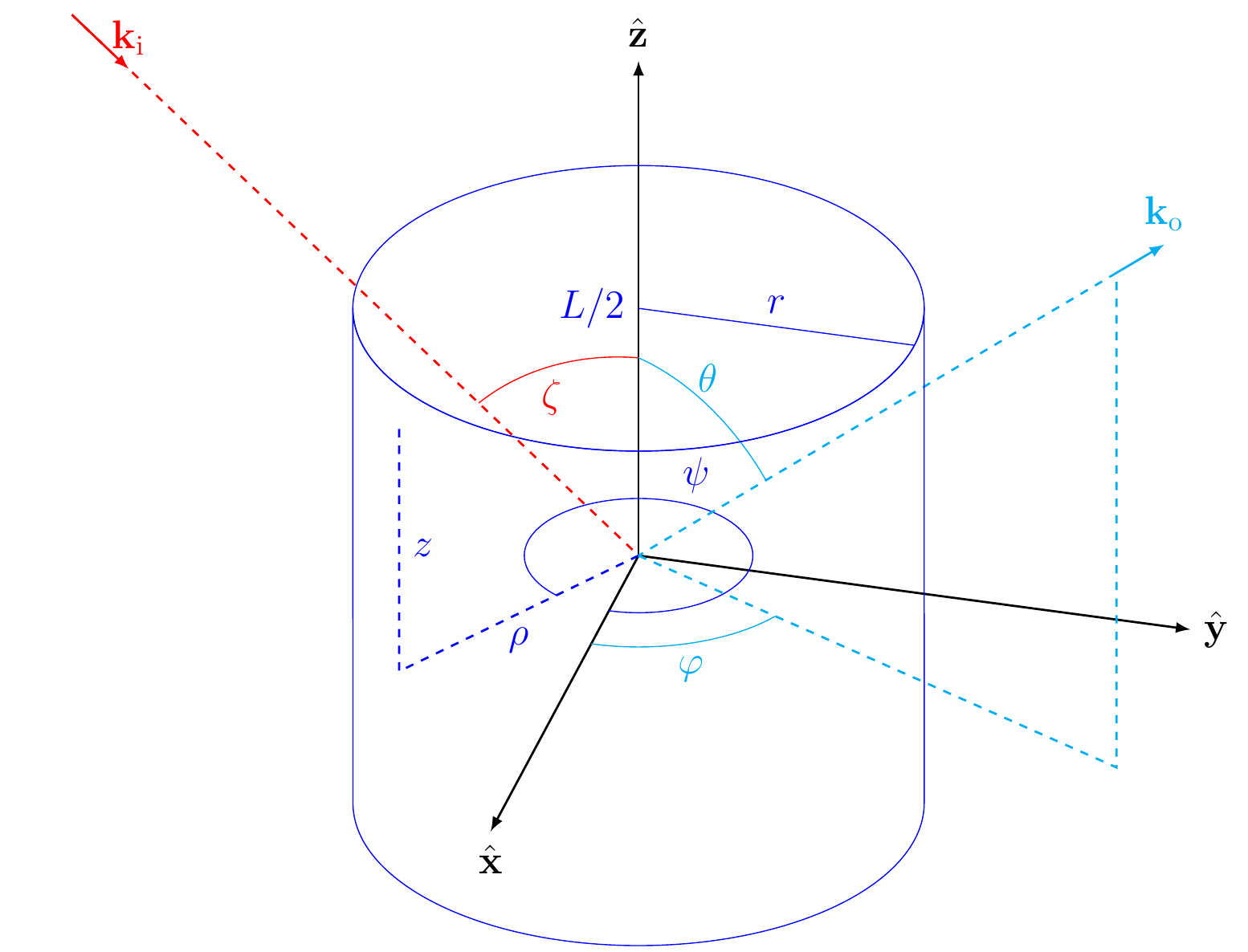}
    \caption{Coordinates used in the derivation of the Rayleigh-Gans-Debye scattering cross-section of a finite cylinder.}
    \label{fig:cylinder}
\end{figure*}

\end{document}